%% file: main.tex
\documentclass[conference,10pt,a4paper]{IEEEtran}
\input{_header.tex}

\usepackage{microtype}
\doubleblindfalse
\hypersetup{
	pdftitle={The MQT Compiler Collection: A Blueprint for a Future-Proof Quantum-Classical Compilation Framework},
	pdfsubject={2026 Design, Automation and Test in Europe Conference (DATE)},
	pdfauthor={Lukas Burgholzer, Daniel Haag, Yannick Stade, Damian Rovara, Patrick Hopf, Robert Wille}
}

\usepackage{flushend}

\begin{document}

\input{_header-document.tex}

\title{The MQT Compiler Collection\\\vspace*{-1mm}{\LARGE A Blueprint for a Future-Proof Quantum-Classical Compilation Framework}\vspace*{-5mm}}

\author{
\IEEEauthorblockN{
Lukas Burgholzer\IEEEauthorrefmark{1}\IEEEauthorrefmark{2},
Daniel Haag\IEEEauthorrefmark{1},
Yannick Stade\IEEEauthorrefmark{1},
Damian Rovara\IEEEauthorrefmark{1},
Patrick Hopf\IEEEauthorrefmark{1}\IEEEauthorrefmark{2},
Robert Wille\IEEEauthorrefmark{1}\IEEEauthorrefmark{2}
}
\IEEEauthorblockA{\IEEEauthorrefmark{1}Chair for Design Automation, Technical University of Munich, Germany}
\IEEEauthorblockA{\IEEEauthorrefmark{2}Munich Quantum Software Company (MQSC), Garching bei M\"{u}nchen, Germany}
\IEEEauthorblockA{\{lukas.burgholzer, daniel.haag, yannick.stade, damian.rovara, patrick.hopf, robert.wille\}@tum.de}
\href{https://www.cda.cit.tum.de/research/quantum}{cda.cit.tum.de/research/quantum}\vspace*{-5mm}
}

\maketitle

\begin{abstract}
As the capabilities of quantum computing hardware continue to rise, algorithms that exploit them are becoming increasingly complex.
These developments increase the need for sophisticated compilation frameworks that translate high-level algorithms into executable code.
In the past, most solutions were built with a \emph{quantum-first} approach and handled mostly pure quantum programs without classical elements such as structured control flow.
However, developments in quantum algorithms, error correction, and optimization, as well as the integration into high-performance computing (HPC) environments, depend on such classical elements.
As quantum-first approaches increasingly struggle to handle these concepts, \emph{classical-first} approaches are becoming a promising alternative.
In this work, we present the \emph{MQT Compiler Collection}, a blueprint for a future-proof \emph{quantum-classical} compilation framework built on the \emph{Multi-Level Intermediate Representation} (MLIR).
After years of experience with the quantum-first approach and its shortcomings, we propose a framework that embraces core MLIR concepts to support the \emph{full compilation pipeline} from high-level algorithms to hardware-specific instructions.
The proposed architecture is designed from the ground up to support complex optimizations beyond, e.g., simple gate cancellation.
It is publicly available at \href{https://github.com/munich-quantum-toolkit/core}{github.com/munich-quantum-toolkit/core}.
\end{abstract}

\input{sections/introduction.tex}

\input{sections/background.tex}

\input{sections/motivation.tex}

\input{sections/four.tex}

\input{sections/conclusion.tex}

\vfill

\input{sections/acknowledgments}

\clearpage

\printbibliography

\end{document}

%% file: _header.tex
\usepackage{graphicx}
\usepackage{tabularx}

\usepackage[T1]{fontenc}
\usepackage[utf8]{inputenc}
\usepackage[english]{babel}

\usepackage{csquotes}
\usepackage{nicefrac}
\usepackage{afterpage}
\usepackage{needspace}
\usepackage[textsize=scriptsize]{todonotes}
\usepackage{booktabs}
\usepackage{tikz}
\usepackage{quantikz}
\usetikzlibrary{arrows,positioning} 
\usetikzlibrary{shapes.geometric, positioning}
\usetikzlibrary{fit, quotes}
\usetikzlibrary{shapes.geometric, positioning}
\usetikzlibrary{automata,arrows}
\usepackage{hyperref}
\usepackage{soul}
\usepackage{amssymb}
\usepackage{csvsimple}
\usepackage{amsthm}
\usepackage{amsmath}
\usepackage{lipsum}
\usepackage{varwidth}
\usepackage{utfsym}
\usepackage{fontawesome5}
\usepackage{multicol}
\usepackage{listings}
\usepackage{adjustbox}
\usepackage{float}
\usepackage{placeins}
\usepackage{yquant}
\usepackage{pgfplots}
\pgfplotsset{compat=1.18}
\usepackage[detect-all=true]{siunitx}
\usepackage{etoolbox}
\usepackage{bbm}
\usepackage{balance}
\usepackage{changes}
\usepackage{makecell}
\usepackage{algorithm}
\usepackage{algpseudocodex}

\robustify\bfseries

\makeatletter
\let\MYcaption\@makecaption
\makeatother
\usepackage[font=footnotesize]{subcaption}
\makeatletter
\let\@makecaption\MYcaption
\makeatother

\newtheorem{example}{Example}

\usepackage[style=ieee, maxnames=6, minnames=3, date=year, doi=true, isbn=false, url=false, backend=biber, sortcites]{biblatex}
\AtEveryBibitem{
  \clearname{editor}
  \clearfield{series}
  \clearfield{isbn}
  \clearfield{issn}
  \clearfield{volume}
  \clearfield{number}
  \clearfield{pages}
}

\newif\ifdoubleblind

\makeatletter
\newcommand{\redacted}[2][]{
  \ifdoubleblind
    \if\relax\detokenize{#1}\relax
      [redacted for double-blind review]%
    \else
      #1%
    \fi
  \else
    #2%
  \fi
}

\makeatother

\newfloat{lstfloat}{htbp}{lop}
\floatname{lstfloat}{\footnotesize Listing}

%% file: _header-document.tex
% Autoref names
\renewcommand*{\figureautorefname}{Fig.}
\renewcommand*{\sectionautorefname}{Section}
\renewcommand*{\subsectionautorefname}{Section}
\def\exampleautorefname{Example}

%% file: sections/introduction.tex
\section{Introduction}
\label{sec:introduction}

\begin{figure}[htbp]
    \centering
    \includegraphics[width=0.85\linewidth]{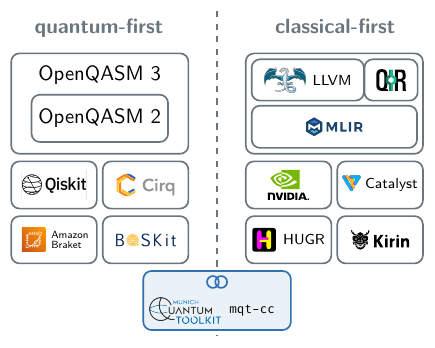}
    \caption{The \emph{MQT Compiler Collection} (\texttt{mqt-cc}) combines quantum and classical concepts to build a future-proof compilation framework.}
    \label{fig:quantum_classical}
\end{figure}

With the rapid growth of quantum software tools and the increasing capabilities of quantum hardware, robust compilation has become a critical bottleneck in the quantum computing stack~\cite{mqss}.
Compilation frameworks serve as the bridge between a diverse ecosystem of high-level programming languages and a heterogeneous zoo of hardware platforms---ranging from superconducting circuits, trapped ions, and neutral atoms to photonics and beyond.
Historically, this bridge has been built using a \emph{quantum-first} approach (see the left half of \autoref{fig:quantum_classical}).
Early \emph{intermediate representations}~(IRs), such as OpenQASM~2, established a standard for allocating qubits and defining gate sequences, effectively serving as the assembly language of early quantum computing.

However, quantum computing is inherently hybrid, relying on classical computing for control, error correction, and optimization, particularly as quantum computers are integrated into classical high-performance computing~(HPC) environments~\cite{mqss,mansfieldPracticalExperiencesIntegrating2025,SHEHATA2026107980,humbleQuantumComputersHighPerformance2021}.
Features taken for granted in classical computing, such as structured control flow, variable scoping, and general-purpose logic, are often entirely absent from quantum-first IRs or added only as ad-hoc extensions.
Although newer standards such as OpenQASM~3~\cite{crossOpenQASMBroaderDeeper2022} aim to address these gaps, many established tools, including Qiskit~\cite{qiskit}, Cirq~\cite{cirq}, Amazon Braket~\cite{braket}, BQSKit~\cite{bqskit}, and the Munich Quantum Toolkit (MQT;~\cite{mqt}), remain constrained by legacy architectures that treat classical computing as a second-class citizen or a mere addition.

In contrast, \emph{classical-first} approaches (see the right half of \autoref{fig:quantum_classical}) take a fundamentally different path.
Instead of reinventing the wheel, they build on decades of classical compiler development and propose to extend established frameworks with quantum concepts.
This approach views the quantum computer as an accelerator---similar to a GPU---integrated into a larger classical computing environment~\cite{kayaSoftwarePlatformSupport2024,sitdikov2025quantum}.
The \emph{Quantum Intermediate Representation} (QIR;~\cite{qir}) marked a significant shift in this direction by extending the widely used LLVM~\cite{llvm} framework for quantum computing.
However, LLVM usually operates on a relatively low level of abstraction.
Its IR is designed for instruction-level optimization and code generation, which makes it challenging to represent and optimize high-level quantum algorithms that might, for example, contain structured control flow.
This gap has led to the rising popularity of the \emph{Multi-Level Intermediate Representation}~(MLIR;~\cite{mlir}).

MLIR addresses these limitations by enabling the definition of custom dialects at various levels of abstraction.
It allows for the seamless composition of compilation passes within a shared infrastructure, promoting modularity, reuse, and interoperability.
However, adopting MLIR poses substantial challenges for tools rooted in the quantum-first paradigm, as it requires moving beyond Python-based rapid prototyping toward disciplined C++ software engineering with a strong focus on stability and maintainability.

In this work, we present the \emph{MQT Compiler Collection} as a blueprint implementation of a future-proof \emph{quantum-classical} compilation framework built on MLIR.
Drawing on our domain expertise from the quantum-first era, we embrace and combine the best of both worlds.
Building on the principles of MLIR, the proposed implementation supports the \emph{full compilation pipeline} from high-level algorithms to hardware-specific instructions.

The remainder of this paper is organized as follows.
\autoref{sec:background} provides the necessary background on MLIR and discusses related work.
In \autoref{sec:motivation}, we motivate our design choices and outline the core principles of the proposed architecture.
\autoref{sec:implementation} details the technical implementation of the \emph{MQT Compiler Collection}.
\autoref{sec:conclusion} concludes the paper.

%% file: sections/background.tex
\section{Background and Related Work}
\label{sec:background}

To ensure that this article is self-contained, we briefly review the relevant aspects of MLIR.
Subsequently, we discuss related approaches and how they address the challenge of defining an infrastructure for quantum compilation.

\subsection{Multi-Level Intermediate Representation}
\label{ssec:mlir}

MLIR addresses the fragmentation of (classical) compilation infrastructure by providing a unified, extensible framework that supports multiple levels of abstraction.
At its core is the concept of \emph{dialects}.
Much like natural languages, each dialect defines its own vocabulary in the form of operations, types, and attributes.
Crucially, these dialects can be mixed within a single program, allowing different concepts to be expressed side by side.
For instance, the \texttt{scf} dialect provides concepts for representing structured control flow (like loops and conditionals), while the \texttt{arith} dialect enables the description of arithmetic operations.
When implementing a new dialect for a specific domain---such as quantum computing---developers can leverage these existing dialects without having to reimplement common functionality.

To make these concepts more graspable, let us have a look at one of the toy examples from the MLIR documentation.

\begin{example}
    The snippet below defines a tensor, transposes it, computes its product with itself, and then prints the result.
    \texttt{toy.constant}, \texttt{toy.transpose}, \texttt{toy.mul}, and \texttt{toy.print} are \emph{operations}, and \texttt{\%0}, \texttt{\%1}, and \texttt{\%2} are the \emph{values} returned by the operations.
    The keyword \texttt{dense} signifies the built-in \emph{attribute} for multi-dimensional arrays, whereas \texttt{tensor} is a built-in \emph{type}.
    
    \begin{center}
        \includegraphics[width=0.85\linewidth]{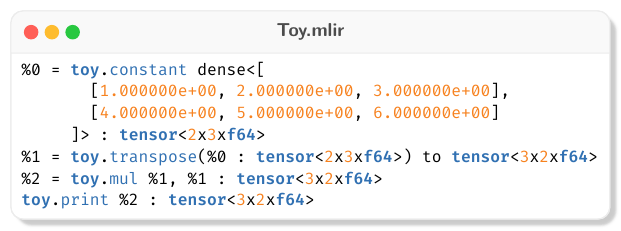}
    \end{center}
\end{example}

The \emph{multi-level} nature of the MLIR framework allows for a progressive compilation strategy.
A program can start at a high level of abstraction, closely matching the source language or algorithm, and be successively \enquote{lowered} to more concrete representations.
This process is facilitated by the \emph{conversion} framework, which handles the transformation of operations
from one dialect to another.
During lowering, operations from high-level dialects can be gradually replaced by operations from lower-level dialects.
In the quantum context, this means one can, e.g., describe a high-level quantum algorithm and progressively lower it to a program that can actually be executed on some specific quantum(-classical) hardware---all within the same infrastructure.

At each stage of the lowering process, MLIR provides powerful optimization tools to simplify the underlying program representation.
One can define \emph{canonicalization} patterns for operations---rules that perform local simplifications (e.g., removing self-inverse gates).
More complex or global optimizations can be implemented as \emph{transformation passes}.
The framework provides standardized interfaces for these passes, e.g., for traversing and manipulating the IR---making it easier to write robust compiler analyses and transformations.

Furthermore, MLIR includes a flexible \emph{translation} framework for importing and exporting external IRs to and from MLIR, respectively.
This allows importing established representations, such as OpenQASM, into MLIR for optimization, and exporting the results back to other formats or directly to executable code.

\subsection{Related Work}

The shift towards classical-first frameworks has been a major trend in recent years, driven largely by the need for scalable and integrated quantum-classical compilation.
A prominent example is Microsoft's Azure Quantum~\cite{azure-quantum}, which heavily leverages QIR to build a robust ecosystem for quantum development.
Similarly, many other industrial players have begun building their tools with a strong connection to classical compilation infrastructure.

NVIDIA's CUDA-Q~\cite{cudaq} focuses on streamlining hybrid application development by providing a unified programming model for CPUs, GPUs, and QPUs.
Central to its architecture is the Quake dialect, which brings quantum concepts into the MLIR ecosystem.
Xanadu's Catalyst~\cite{catalyst}, inspired by the design of QIRO~\cite{qiro}, also leverages MLIR to enable just-in-time compilation of hybrid quantum programs.
It captures Python-based programs and translates them into a representation that can be optimized and executed efficiently.

Other approaches draw inspiration from MLIR's design philosophy without directly building upon it.
Quantinuum's HUGR~\cite{hugr} introduces a hierarchical, graph-based IR for mixed quantum-classical programs.
It is implemented in Rust and emphasizes a unified representation for various stages of compilation.
QuEra's Kirin~\cite{kirin} follows a similar path, aiming to provide a flexible compilation infrastructure for embedded domain-specific languages.
These efforts collectively highlight the growing consensus that modular, multi-level IRs are essential for the future of quantum compilation.

%% file: sections/motivation.tex
\section{Motivation}
\label{sec:motivation}

Building upon insights from both quantum-first and classical-first compilation efforts, we present the \emph{MQT Compiler Collection}---a comprehensive quantum-classical compilation framework built on MLIR that supports the full pipeline from high-level algorithms to hardware-specific instructions.
Rather than viewing these paradigms as competing approaches, we seek to combine their respective strengths: the domain expertise and optimization techniques cultivated in quantum-first tools with the robust infrastructure and proven compiler methodology of classical-first frameworks.
By embracing and integrating concepts from both worlds, the MQT Compiler Collection serves as a blueprint for future-proof compilation infrastructure.
We argue that the core principles of MLIR---modularity, specialized dialects, and progressive lowering---provide an ideal foundation for achieving this synthesis.

\subsection{The MLIR Opportunity}

The need for a compilation infrastructure that handles both quantum and classical aspects is evident.
However, adopting MLIR is not just about using a specific software framework; it is about adopting a \emph{mindset} of progressive lowering and modular abstraction.
Rather than building a monolithic compiler that compiles high-level algorithms into hardware pulses in a single step, the MLIR philosophy encourages breaking the problem down into smaller, more manageable steps.

While one could implement such a layered architecture from scratch, MLIR offers a proven infrastructure for defining dialects, conversions, transformation passes, and more.
This is particularly valuable for the quantum community, as it avoids the need to build and maintain custom implementations for common compiler tasks.
Since quantum computers are increasingly viewed not as standalone devices but as specialized accelerators working in tandem with CPUs and GPUs (e.g., in HPC environments), a framework inherently designed for such heterogeneity is a compelling choice.
By treating quantum operations as just another dialect within the broader compilation landscape, we can reuse existing analyses and optimizations for the classical parts of the program, such as loop optimizations and constant folding.
This is why we have opted to implement the proposed quantum compilation framework in MLIR, fully embracing the classical-first mindset by adopting established MLIR patterns and techniques.

\subsection{Design Philosophy}
\label{ssec:proposed-architecture}

After years of experience with the quantum-first approach, we aim to apply our expertise to build a future-proof quantum-classical compilation framework.
The proposed framework retains the domain-specific capabilities of our previous efforts~(see, e.g., Refs.~\cite{mqt}~and~\cite{burgholzer2025MQTCore}) while leveraging MLIR's robust infrastructure to overcome previous limitations.
Our goal is to provide a blueprint implementation that supports the full compilation pipeline from high-level algorithms to hardware-specific instructions.
The proposed architecture is built to be extended and adapted as the field evolves.
To achieve this, our design philosophy is guided by several key principles:

\begin{itemize}
    \item \textbf{Full-Stack Support}:
    We explicitly target the full compilation stack and support the full compilation pipeline from high-level algorithms to hardware-specific instructions.
    Modifiers and custom gate definitions ensure extensibility and flexibility throughout the lowering process. 
    \item \textbf{Next-Generation Programs}:
    We look beyond static circuits to support structured quantum programs with control flow, dynamic circuits, and hybrid quantum-classical workflows.
    \item \textbf{Comprehensive Optimization}:
    The proposed architecture is designed to support a fully-fledged optimization framework.
    We aim beyond simple gate cancellation and peephole optimizations to enable complex passes such as high-level synthesis, placement and routing, and non-local optimizations.
    \item \textbf{Rigorous MLIR Adoption}:
    We do not just use MLIR as a mean to an end; we embrace its methodology.
    This means not reinventing the wheel, but leveraging existing dialects (e.g., \texttt{arith} for arithmetic operations) as well as built-in features and data structures.
    \item \textbf{Modern Standards}:
    We adopt modern standards to ensure longevity and compatibility.
    We build against the latest stable releases of LLVM/MLIR, embrace QIR~2.0 (with opaque pointer support), and support OpenQASM~3.x.
\end{itemize}

To this end, the architecture leverages two distinct dialects, each designed around a fundamental programming paradigm:

\begin{itemize}
    \item \textbf{QC (Quantum Circuit)}:
    The QC dialect serves as a natural interface for input/output and hardware mapping.
    It utilizes \emph{imperative} concepts and \emph{reference} semantics to mirror the physical reality of quantum devices.
    \item \textbf{QCO (Quantum Circuit Optimization)}:
    The QCO dialect adopts a (first-order) \emph{functional} approach with \emph{value semantics}, providing an explicit representation ideally suited for complex optimizations and transformations.
\end{itemize}

\begin{figure}[b]
    \centering
    \includegraphics[width=0.95\linewidth]{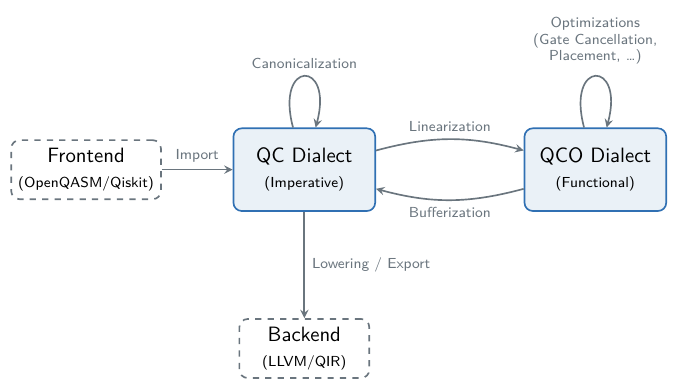}
    \caption{The MQT Compiler Collection pipeline.}
    \label{fig:pipeline}
\end{figure}

By supporting both representations within the same framework, we can leverage the strengths of each model without compromise.
As illustrated in \autoref{fig:pipeline}, this enables a seamless flow from high-level language descriptions to hardware-executable code, while providing opportunities for optimization at different levels of abstraction.

The compilation flow starts with a circuit description in a language or framework such as OpenQASM or Qiskit.
This input is imported into the QC dialect, where the program structure is preserved in an imperative form.
From there, the representation is converted to the QCO dialect by converting qubit references into quantum state values (also called \emph{linearization}).
In this functional representation, various optimization passes are applied to the program.
Canonicalization and gate cancellation prune redundant operations, while backend-specific passes---such as a mapping algorithm for superconducting qubit architectures---adapt the circuit to device constraints.
These passes are merely examples of the transformations enabled by the framework; the infrastructure makes it easy to add further optimizations as needed.
Once optimized, the circuit is converted back to the QC dialect by reconstructing qubit references from the data flow.
Finally, the result is lowered to low-level representations, such as LLVM IR and QIR, enabling execution on simulators or physical quantum hardware~\cite{stade2025TowardsQIR}.
The infrastructure is easily extensible, as demonstrated by an integration of PennyLane via an MLIR plugin~\cite{Hopf_Integrating_Quantum_Software_2026}.

%% file: sections/four.tex
\section{The MQT Compiler Collection}
\label{sec:implementation}

In this section, we present the technical details of the \emph{MQT Compiler Collection}, an open-source implementation of the architecture proposed in \autoref{ssec:proposed-architecture}.
We detail the design of the two core dialects---QC and QCO---and highlight how their imperative and functional natures complement each other to facilitate both interoperability and optimization.
Due to space limitations, we focus on key concepts and representative examples to illustrate the core ideas.
The full implementation is available at \href{https://github.com/munich-quantum-toolkit/core}{github.com/munich-quantum-toolkit/core}.

\subsection{The QC Dialect: An Imperative Interface}

The QC dialect is designed to act as the primary interface between external quantum programming languages and the compiler.
Most established languages, such as OpenQASM, and frameworks, such as Qiskit, follow an \emph{imperative} programming model,
where qubits are treated as mutable resources that are modified in place by sequential operations via side effects.
The QC dialect mirrors these \emph{reference semantics} in its \texttt{!qc.qubit} type.
It defines a comprehensive set of operations:
\begin{itemize}
    \item \textbf{Resource Management}: \texttt{qc.alloc} and \texttt{qc.dealloc} for allocating and deallocating qubits as well as support for qubit registers.
    \item \textbf{Standard Gate Library}: A comprehensive library of well-known gates (e.g., \texttt{qc.h}, \texttt{qc.x}, \texttt{qc.rz}).
    \item \textbf{Non-Unitary Operations}: \texttt{qc.measure} for extracting information and \texttt{qc.reset} for re-initializing qubits.
    \item \textbf{Modifiers}: First-class support for controlled (\texttt{qc.ctrl}), inverse (\texttt{qc.inv}), and power (\texttt{qc.pow}) modifiers.
    \item \textbf{Structure}: Operations for grouping instructions and defining custom gates.
\end{itemize}

\begin{figure}[htbp]
    \centering
    \includegraphics[width=0.9\linewidth]{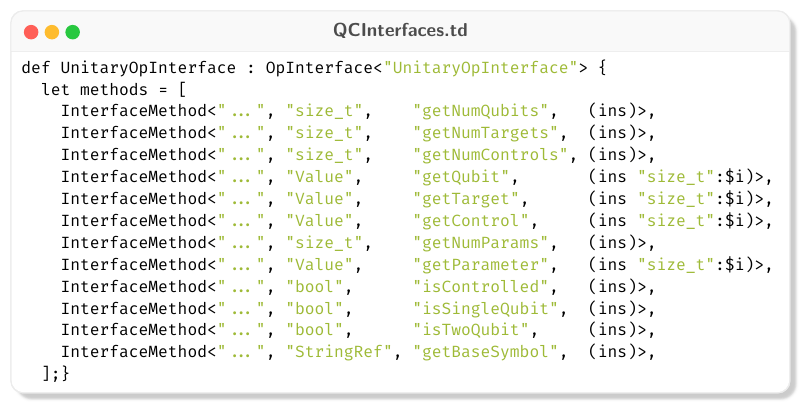}
    \caption{Abbreviated TableGen definition of the \texttt{UnitaryOpInterface}.}
    \label{fig:qc-interfaces}
\end{figure}

All unitary operations, including standard as well as custom gates and gates wrapped in modifiers, implement a common \texttt{UnitaryOpInterface} that provides a unified API for introspection and manipulation (see \autoref{fig:qc-interfaces}).

\begin{example}
    \label{ex:qc-bell}
    Consider the creation of a Bell pair.
    In the QC dialect, the reference semantics are evident: we allocate two qubits and apply operations to them sequentially.
    \texttt{\%q0} and \texttt{\%q1} represent the qubit references themselves, which are reused throughout the program.
    The data flow is \emph{implicit}---defined only by the program order.
    \begin{center}
        \includegraphics[width=0.85\linewidth]{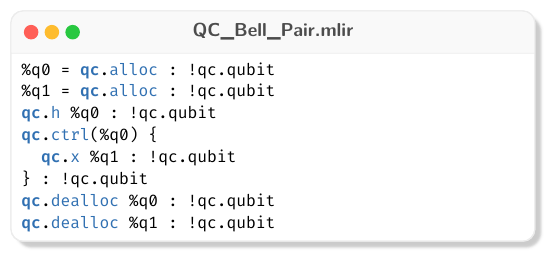}
    \end{center}
\end{example}

This representation is ideal for interacting with other tools and for the final mapping to hardware, but the implicit data flow can obscure dependencies, making advanced optimizations challenging.
That said, simple cleanup tasks can already be implemented in the QC dialect to ensure that the input is not unnecessarily bloated.

\begin{example}
    A common cleanup task is removing pairs of allocation and deallocation operations that have no intervening computations.
    In the QC dialect, this requires verifying that the allocated qubit has no uses other than its deallocation.
    The optimization pattern matches a \texttt{DeallocOp}, looks up the defining \texttt{AllocOp}, and checks if the qubit is used only once.

    \begin{center}
        \includegraphics[width=0.9\linewidth]{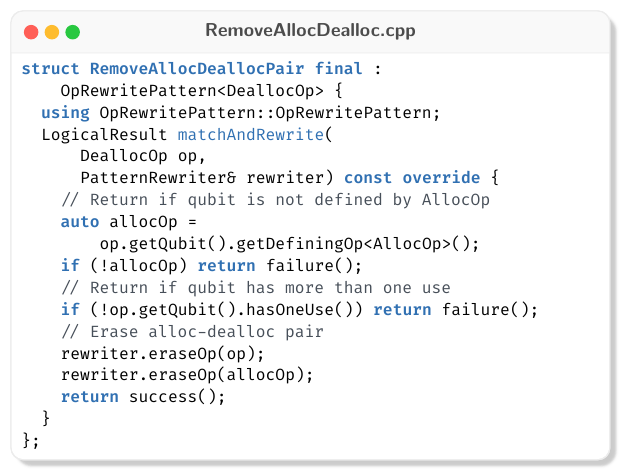}
    \end{center}%
    Applying this pattern simplifies the IR as shown below:

    \begin{center}
        \includegraphics[width=0.9\linewidth]{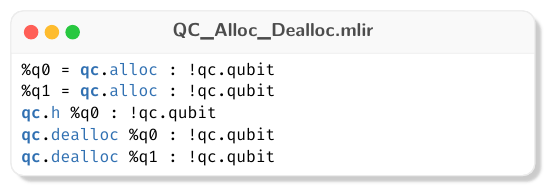}
    \end{center}

    \begin{center}
        \includegraphics[width=0.9\linewidth]{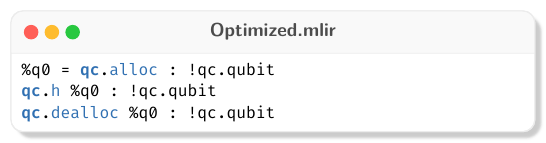}
    \end{center}
\end{example}

\subsection{The QCO Dialect: A Functional Optimization Graph}

To enable more complex circuit transformations, we have designed the QCO dialect.
In contrast to QC, QCO adopts \emph{value semantics} that are typical for \emph{functional} programming languages.
Here, operations do not modify qubits in place.
Instead, they consume input quantum states and produce new, updated output states---effectively modeling qubits as linear types.
This transforms the circuit into a dataflow graph where dependencies are \emph{explicit}.

This structure is conceptually identical to the directed acyclic graphs (DAGs) that frameworks like Qiskit build internally to analyze and optimize circuits.
However, while those frameworks have generally implemented custom DAG data structures and algorithms from scratch, the QCO dialect directly uses the use-def chains of MLIR values to represent this graph.
In other words, the QCO dialect natively embeds the DAG representation within the IR itself.
This design choice allows us to leverage MLIR's built-in analysis and transformation capabilities directly on the dataflow graph.

\begin{example}
    Revisiting the Bell pair circuit in the QCO dialect illustrates the explicit data flow.
    The \texttt{qco.h} operation consumes the initial state of the first qubit (\texttt{\%q0\_0}) and produces a new state (\texttt{\%q0\_1}).
    The subsequent controlled operation explicitly depends on \texttt{\%q0\_1} as its control input.
    Because every operation produces new values, the dependency graph is encoded directly in the program structure.

    \begin{center}
        \includegraphics[width=0.9\linewidth]{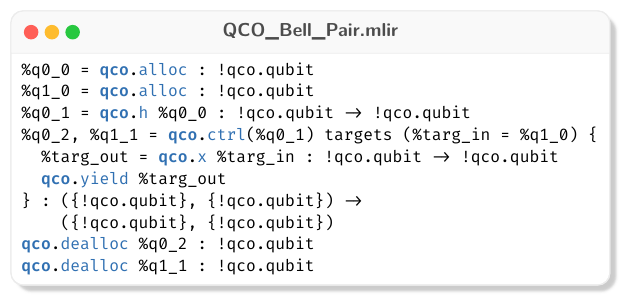}
    \end{center}
\end{example}

\subsection{Optimizations Made Easy}

The explicit data flow of the QCO dialect significantly simplifies the implementation of optimization passes.
Many circuit optimizations can be expressed as simple pattern-matching rules (canonicalization patterns) on the dataflow graph.
While we have already shown an example of such a pattern in the QC dialect, it was only reasonable to implement it there since it was enough to check the \emph{number} of uses.
Slightly more complex optimizations, such as gate cancellation, immediately profit from the value semantics of QCO.
Because dependencies are explicit, detecting if two operations act on the same qubit in sequence is trivial---one simply checks if the input of the second operation is the output of the first.

\begin{example}
    \label{ex:h-canonicalization}
    Hermitian gates like the Hadamard gate satisfy $H^\dagger H = I$, i.e., they are self-inverse.
    Thus, any two adjacent Hadamard gates cancel each other out.
    Implementing this check in QCO is concise because the adjacency is guaranteed by the linear data flow.
    The following shows a generalized pattern for cancelling any two types of consecutive inverse operations:

    \begin{center}
        \includegraphics[width=0.9\linewidth]{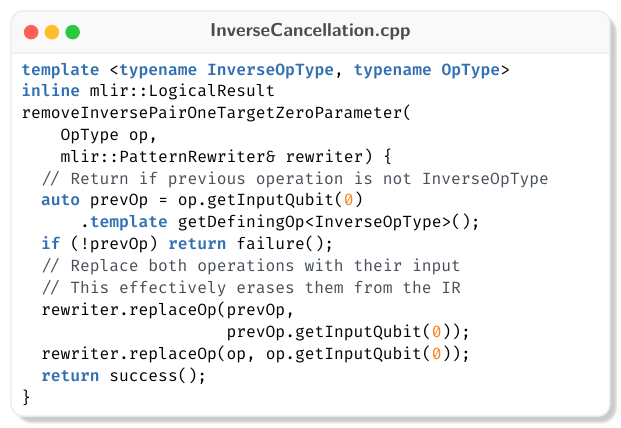}
    \end{center}%
    Applying this pattern simplifies the IR as shown below:

    \begin{center}
        \includegraphics[width=0.9\linewidth]{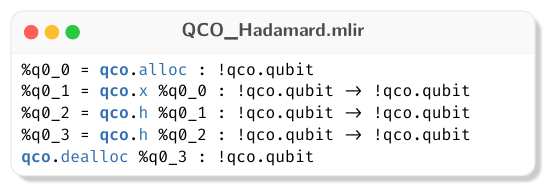}
    \end{center}

    \begin{center}
        \includegraphics[width=0.9\linewidth]{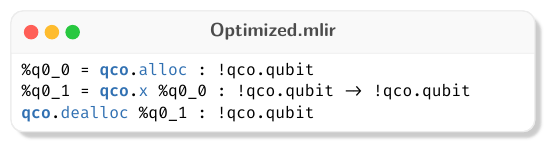}
    \end{center}
\end{example}

Beyond local canonicalization patterns, the proposed framework supports sophisticated transformations.
We have ported the QMAP mapping algorithm~\cite{zulehnerEfficientMethodologyMapping2019} to MLIR as a representative example of a complex, backend-aware pass.
This pass maps quantum circuits to architectures with limited connectivity.
Building on the existing MLIR infrastructure enabled us to support routing for circuits with structured control flow---a feature that was not possible in the previous open-source implementation without significant engineering effort~\cite{wille2023qmap}.

\subsection{Bridging the Paradigms}

To connect the imperative world of QC with the functional world of QCO, we provide robust bidirectional conversion passes.
These passes automate the translation between reference and value semantics.
Converting from QC to QCO involves \enquote{linearization} or building the dataflow graph via tracking the latest value of each qubit and introducing new values for each operation.
Conversely, converting back to QC involves \enquote{bufferization} or reconstructing qubit references from the data flow by ensuring that each qubit value is replaced with the appropriate reference.

\begin{example}
    The code below shows the core logic for converting a single-qubit gate from QC to QCO.
    The compiler tracks the current value for each qubit using \texttt{getLatestQCOQubit}.
    It creates the corresponding QCO operation using the latest value and updates the state with the new output value using \texttt{updateQCOQubitMap}.

    \begin{center}
        \includegraphics[width=0.85\linewidth]{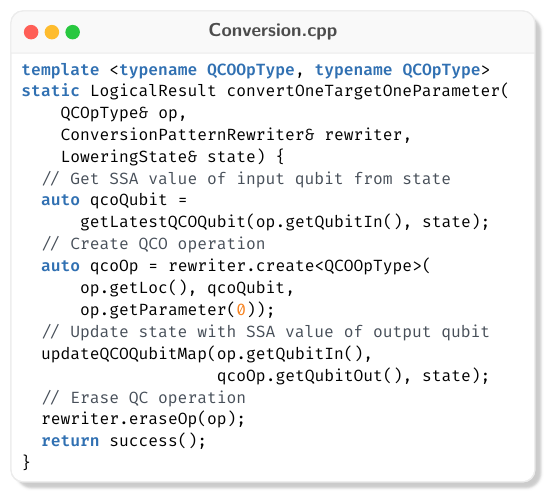}
    \end{center}
\end{example}

Conversions for more complex constructs, such as the control modifier, require delicate threading of the qubit values to ensure the linearity constraints are upheld.
However, the modular nature of MLIR's conversion framework makes it straightforward to implement and maintain these transformations.

\subsection{Structured Control Flow}

As stressed in \autoref{ssec:proposed-architecture}, any future-proof quantum-classical compilation framework must support structured quantum programs.
That is, it must support programs that do not only consist of a static sequence of quantum operations but also include classical control flow that may, additionally, depend on measurement outcomes.

MLIR's built-in \texttt{scf} dialect provides operations for defining loops (such as \texttt{scf.for} and \texttt{scf.while}) and conditionals (such as \texttt{scf.if}), along with predefined canonicalization patterns.
These operations can be seamlessly integrated into an IR utilizing the QC dialect.
Let us illustrate this with a simple example.

\begin{example}
    The following program demonstrates a native reset operation using structured control flow.
    A qubit is prepared in a superposition state via a Hadamard gate.
    The qubit is subsequently measured, yielding a classical bit \( c \in \{ 0, 1 \} \).
    If the measurement outcome is \( c = 1 \), the qubit is flipped using an X gate; otherwise, no operation is performed.
    As a result, the qubit is always reset to the \(\lvert 0 \rangle\) state.

    \begin{center}
        \includegraphics[width=0.85\linewidth]{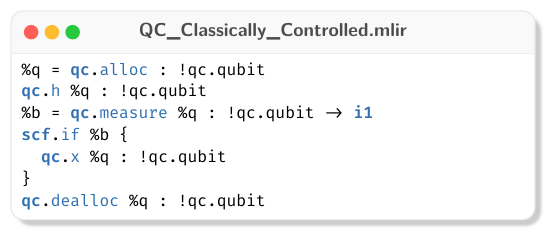}
    \end{center}
\end{example}

This goes to show how natural it is to express dynamic quantum circuits using structured control flow in the proposed framework.
Moreover, the integration with MLIR's \texttt{scf} dialect allows us to directly leverage existing optimizations for these constructs.
For instance, loop unrolling and conditional simplifications can be applied without any additional effort.

%% file: sections/conclusion.tex
\section{Conclusion}
\label{sec:conclusion}

In this work, we have presented the \emph{MQT Compiler Collection}---a blueprint implementation that demonstrates how to build future-proof quantum-classical compilation infrastructure on MLIR.
By combining insights from both quantum-first and classical-first paradigms, we have shown that it is possible to support the full compilation pipeline from high-level algorithms to hardware-specific instructions within a unified, modular framework.
At its core, the proposed dual-dialect architecture exemplifies the power of MLIR's design philosophy: the imperative QC dialect provides a natural interface for external tools and hardware, while the functional QCO dialect enables sophisticated optimizations through value semantics.

Beyond the technical contributions, this work serves as a community resource.
Developed entirely as open-source software, the MQT Compiler Collection offers a concrete starting point for researchers and practitioners looking to adopt MLIR for quantum compilation or to extend the framework with new capabilities.
We hope that by sharing both our implementation and the lessons learned from bridging quantum-first expertise with classical-first infrastructure, we can accelerate the development of robust, interoperable compilation frameworks across the quantum computing ecosystem.
The full implementation is publicly available as part of MQT Core~\cite{burgholzer2025MQTCore} at \href{https://github.com/munich-quantum-toolkit/core}{github.com/munich-quantum-toolkit/core}.

%% file: sections/acknowledgments.tex
\section*{Acknowledgments}
{\footnotesize
The authors acknowledge funding from the European Research Council (ERC) under the European Union’s Horizon 2020 research and innovation program grant agreement No. 101001318 and No. 101114305 (“MILLENION-SGA1” EU Project), and the Munich Quantum Valley, which is supported by the Bavarian state government with funds from the Hightech Agenda Bayern Plus. Furthermore, this work was supported by the BMFTR under grant numbers 13N17298 (SYNQ) and 01MQ25001I (FullStaQD), the Deutsche Forschungsgemeinschaft (DFG, German Research Foundation) under grant numbers 563402549 and 563436708, and the Austrian Research Promotion Agency (FFG) together with the states of Upper Austria and Tyrol within the COMET module Quantum Algorithm Engineering (FFG) under grant number 923923.
}